\documentclass[ES]{rcftex}

\title{Estudios num{\'e}ricos de Aerofractures en medios porosos}

\titleEN{Numerical Studies of Aerofractures in Porous Media} 
  \author{Michael J. Niebling\toaff{a,b,c \emailto}, Renaud Toussaint\toaff{b,c,d}, Eirik G. Flekk\o y\toaff{a,d} and Knut J\o rgen M\aa l\o y\toaff{a,d}}
  \affiliations{
\aff Department of Physics, University of Oslo, P.O. Box 1048, 0316 Oslo, Norway. Michael.Niebling@fys.uio.no$^\emailto$
\aff Institut de Physique du Globe de Strasbourg, CNRS 
\aff University of Strasbourg, 5 rue Descartes, 67084 Strasbourg Cedex, France
\aff Centre for Advanced Study at the Norwegian Academy of Science and Letters, Drammensveien 78, 0271 Oslo, Norway}

\abstractES{Durante la compactaci{\'o}n hidraulicamente inducida de una capa granular surgen patrones de fractura. Con simulaciones num{\'e}ricas estudiamos c{\'o}mo estos patrones dependen de las propiedades del gas intersticial asi como de las propiedades del medio poroso. En particular aqu{\'i} se estudia en detalle la relaci{\'o}n entre la velocidad de propagaci{\'o}n de fractura y la presi{\'o}n de inyecci{\'o}n.\\}

\abstractEN{During the hydraulically induced compaction of a granular layer fracture patterns arise. In numerical simulations we study how these patterns depend on the gas properties as well as on the properties of the porous medium. In particular the relation between the speed of fracture propagation and injection pressure is here studied in detail.\\}
\keyW{Hydrofractures, compressible fluid, porous material, coupled grain fluid flow.}
\begin{document}
\maketitle
\section{Introduction}
Stress induced by fluid or gases can cause diverse materials to break and fracture. Such hydraulic fractures are a natural and common phenomenon in the field of volcanism and are artificially initiated to enhance the recovery of natural gas and mineral oil by fracturing the reservoir rock with pressurized fluids. Recently a new perspective on hydrofractures was added with the storage of supercritical $\mathrm{CO}_2$ attracting the interest of an increasing number of researchers. In this respect two scenarios are considered. First it is one option to inject $\mathrm{CO}_2$ into existing hydrofractures, and second the injection of the $\mathrm{CO}_2$ can create additional fractures \cite{Sleipner,Sleipner2}. The typical components for such fractures are a porous material and a compressible gas. Injection of pressurized fluids in a porous material, deforming beyond the elastic limit, has been studied in granular materials in Hele-Shaw cells,\cite{VanDamme,Nat,Huang,airin,airin2,airin3}, with the injection of air or oil in systems with open boundary conditions, and during cyclic loading \cite{An1}. 
It was also studied in systems with a confinement for the grains, prevented from getting out of the cell, which allowed to observe the formation of thin fractures \cite{nieb4}. In this paper \cite{nieb4} it was found and discussed a criterion that the porous media and the fluid need to fulfill to allow the formation of fractures. For this purpose the gas' viscosity was varied. It was further discussed how the shape of the fractures depend on the properties of the porous material and of the injected gas in simple 2 dimensional (2D) numerical simulations. \\
In contrast to the previous article we will not change the properties of the injected gas or the porous material in this present article. Here we explore in particular the effect of the amplitude of the fluid pressure imposed in the source on the fracture morphology. Furthermore, all simulations here will be ran in a regime where fractures are created.
\section{Simulation setup}
\begin{figure}
\includegraphics[width=6cm]{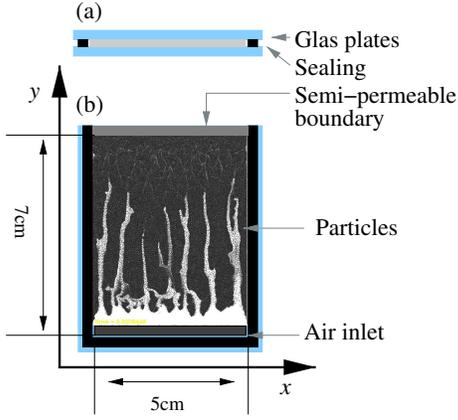}
\caption{\label{Setup} Numerical setup of the system.}
\end{figure}

As shown in Fig. \ref{Setup} the setup consists of a cell with two glass plates separated by 1 mm. The gap between the plates is filled with particles. The empty space between the grains is saturated with a fluid that has the same properties as the fluid that is injected. Consequently, the only two media involved in the dynamics are the grains and the fluid. At the start of the simulations the average solid volume fraction of the grains is $\rho_s^{(0)}=0.42$. This starting solid volume fraction is homogeneous with negligible density fluctuations although the particles are at random positions. The value of $\rho_s^{(0)}=0.42$ is chosen to be less than the possible maximum of $\rho_s^{(\mathrm{max})}=0.60$ to allow compaction of the grains. On the inlet side of the cell the pressure is imposed to $P_I$. Several simulations are performed with $P_I$ ranging from a value of $ P_I=0.5 \cdot 10^5$ Pa to a value of $P_I=2.5 \cdot 10^5$ Pa above the atmospheric pressure of $P_0=1.0 \cdot 10^5$ Pa. On the opposing side to the inlet the cell has an open boundary for the fluid but particles are not able to leave the cell here. In a real experiment, this could be achieved by using a net with a mesh smaller than the particles. The remaining boundaries are completely sealed for both media. In the simulation around 200 000 grains of diameter 140$\pm 10\%$ $\mu$m are involved. Finally, the pressure at the inlet is increased and maintained as a step function in time, at a steep ramp, and particles hardly move before the maximum injection pressure $ P_I$ is reached. 
\section{Theory and Model}
Using a well tested numerical model we have the freedom to explore the parameter space independently. The details of the method can be found in \cite{airin,nieb1,nieb2,nieb3,mcnam,jlv,jlv2,jlv3}, and alternative models can be found in \cite{MAvH1,MAvH2,MAvH3,MAvH4,MAvH5}. The model describes the fluid in terms of a pressure field while the porous medium is modeled by simulating discrete particles.
\section{Theory and Simulations}
\subsection{Dynamics of the gas phase}
The equation for the evolution of the pressure $P = \tilde{P} + P_0$, where $P_0$ is the atmospheric pressure and $\tilde{P}$ the local pressure fluctuations is given by:
\begin{equation}
  \label{airP}
  \phi\left[\frac{\partial P}{\partial t}+{\bf u}\cdot\nabla P\right]=\nabla\cdot\left[ P\frac{\kappa}{\mu_f}\nabla P\right]-P\nabla \cdot {\bf u}.
\end{equation}
In this equation the pressure is described in terms of the local granular velocity {\bf u}, the viscosity $\mu_f$ of the gas, the local porosity $\phi=1-\rho_s$ and the local permeability $\kappa$.
Eq. \ref{airP} is derived from mass conservation of the gas and the granular medium and by assuming a local Darcy law. \\
\subsection {Dynamics of the particles}
For the particles we basically use Newton's second law:
\begin{equation}
\label{force_sp}
 m \frac{dv_p}{dt} = {\bf F_I} + {\bf F_{d}} + {\bf F_a} - \frac{\nabla P}{\rho_n},
\end{equation}\\
with particle velocity $v_p$, particle mass $m$, particle mass density $\rho_m$, cell spacing $h$ and the number density $\rho_n=\rho_s\rho_m/m$. ${\bf F_I}$ are linear inter-particle solid contact forces. ${\bf F_{d}}$ is a viscous damping force during particle collisions.\\
For $F_a$, the interaction with the side plates we assume that the normal stress $P_g^{\bot}$ in the granular packing is proportional to the in-plane stress $P_g^{||}$ by a factor $\lambda$ (Janssen hypothesis). 
Using further a Coulomb friction model we state that the frictional force $F_a$ per particle with the glass plates is proportional to the normal stress by a friction coefficient $\gamma$. With these two assumptions we find an expression for the friction force with the side plates.   
\begin{equation}
\label{Fa}
F_a\leq \gamma S_a ( 2 P_g^{\bot}+ \rho_m g h) = \gamma S_a ( 2 \lambda P_g^{||}+ \rho_m g h).
\end{equation}
$S_a=\pi a^2$ is the contact area of the particles with the plates.
\section{Results}
\begin{figure}
\includegraphics[width=8cm]{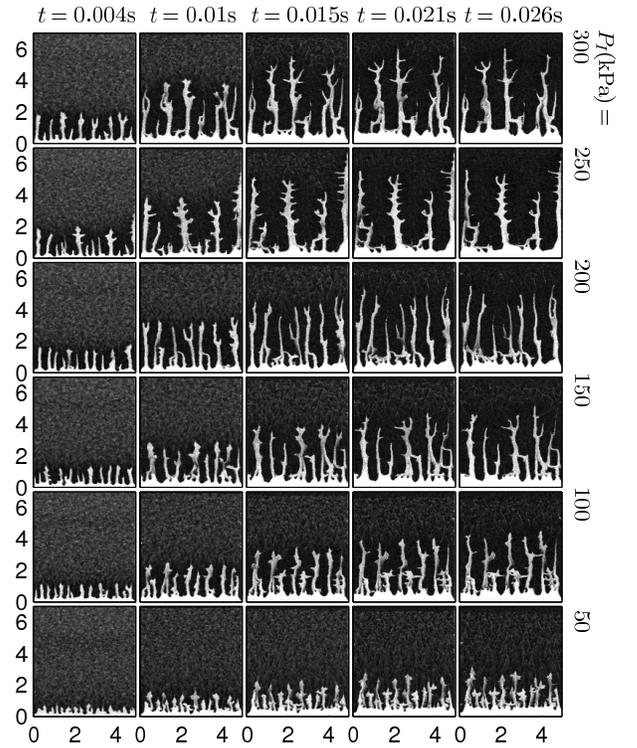}
\caption{\label{density} Snapshots during the simulations of the particle density in the Hele-Shaw cell, displayed for decreasing injection pressure $P_I$ from top to bottom and as a function of time (left to right). Low particle density appears brighter in the snapshots. Under air injection, fractures and fingers of low particle density emerge and propagate. $x$- and $y$-axis units are given in cm. The $y$-axis specifies the distance from the inlet.}
\end{figure}

\begin{figure}
\includegraphics[width=8cm]{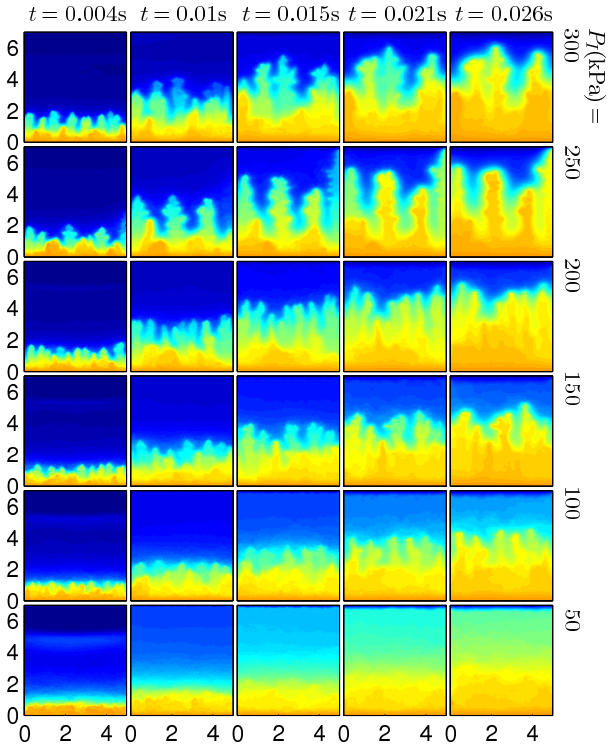}
\caption{\label{pressure} (Color online) The pressure evolution for decreasing injection pressure $P_I$ (top to bottom) and as a function of time (left to right). High pressure appears yellow (brighter) in the snapshots. $x$- and $y$-axis units are given in cm.}
\end{figure}

\begin{figure}
\includegraphics[width=8cm]{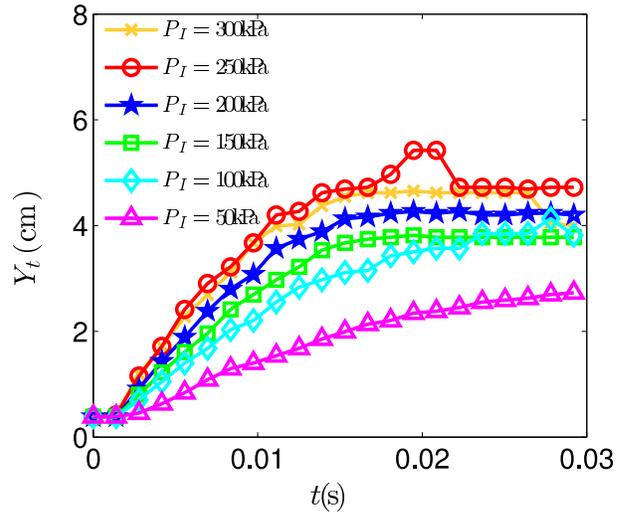}
\caption{\label{finger_position} (Color online) The position of the most advanced finger/fracture as a function of time at different injection pressure $P_I$. The higher the injection pressure $P_I$ the further fingers grow.}
\end{figure}

\begin{figure}
\includegraphics[width=8cm]{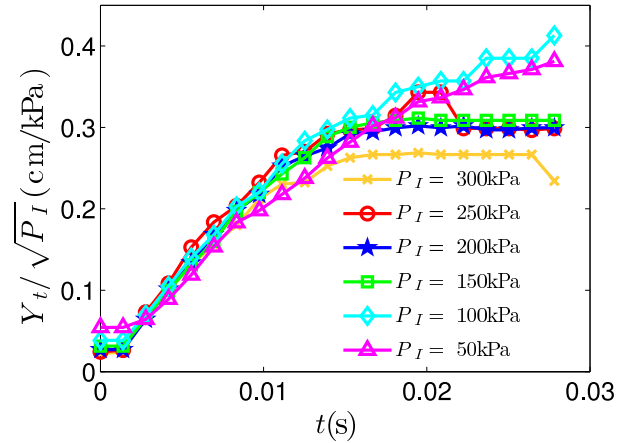}
\caption{\label{finger_position_scaled} (Color online) The position of the most advanced finger/fracture rescaled by the square root of the injection pressure $\sqrt{P_I}$. As a function of time the graphs at different injection pressure $P_I$ collapse onto a single graph.}
\end{figure}

\begin{figure}
\includegraphics[width=8cm]{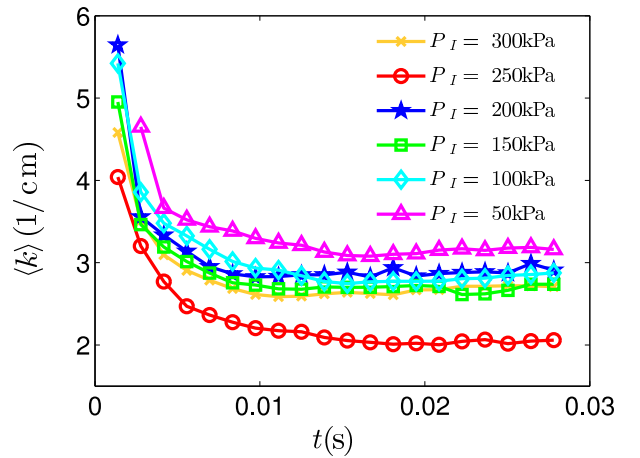}
\caption{\label{finger_k} (Color online) The average of the spatial finger wavenumber in $x$ direction.}
\end{figure}

We ran a set of six simulations for injection pressures of $P_I=(0.5, 1.0, 1.5, 2.0, 2.5, 3,0) \cdot 10^5$ Pa above atmospheric pressure $P_0$, a fluid viscosity of  $\mu_f=18.0$ mPa$\cdot$s and a friction coefficient with the side plates of $\gamma \lambda=4.0$. 
The injected gas is considered as an ideal gas and has the compressibility of air $\beta_T=1/P_0$ at $P_0$. The value of $P_I$ at the inlet is reached very fast and particles start to move shortly after. During this compression of the particles fractures emerge in the granular packing. In Fig. \ref{density} a set of snapshots of the particle density is shown. The snapshots are taken at increasing time from left to right. Each horizontal row of pictures corresponds to one of the six simulations at a different injection pressures. In Fig. \ref{pressure} snapshots of the corresponding pressure field in the cell are displayed. The pressure field is normalized to one to allow a qualitative comparison.
In these plots an apparent feature is the different propagation speed and position of the emerging fractures. A high injection pressure causes the fractures to propagate faster. To quantify this observation we can plot the position of the most advanced finger tip as a function in time. This is done in Fig. \ref{finger_position}. The plot clearly proves the previous observation. Furthermore it turns out that the systematic increase of the propagation speed is also proportional to the square root of the injection pressure. This is checked in Fig. \ref{finger_position_scaled}. Here the rescaling of the fracture tip position by the square root of the injection pressure $\sqrt{P_I}$ results in a collapse of the graphs. The disagreement at the later stages of the simulations in this plot results from the finite size of the system, which allows fractures to grow only up to a certain size. Finally we can state that the fingers grow according to:
\begin{equation}
\label{scaling}
Y_t = \sqrt{P_I} f(t).
\end{equation}
Where $f(t)$ is a function which appears in the plots to be almost linear at early stages of the finger growth for $t<0.01$s.
In Fig. \ref{density} we also observe that the fingers at high injection pressure propagate further into the packing before complete compaction of the grains takes place. This can be also seen in Fig. \ref{finger_position} where the finger position stops growing at longer distances from the inlet the higher the injection speed is.\\
Apart from the finger position, the increase of the injection pressure also affects the shape of the fingers. In Fig. \ref{density} it can be seen that the fingers get more branched and fracture-like at higher injection pressure. At low injection pressure fingers appear to be straighter while increasing the injection pressure, fingers develop more and more branches. At the highest injection pressure of $P_I= 3,0 \cdot 10^5$ Pa the fingers clearly show characteristics of fractures. \\
At high injection pressure $P_I$ the pressure gradients are the largest. When the boundary is deformed the expected changes of the pressure gradients are therefore also higher at high injection pressure than at low injection pressure. At low $P_I$, one expects a lower pressure gradient everywhere, and thus a low effect of seepage forces and a slower deformation. Leading to overall smoother pressure gradients and a more stable front deformation. We thus expect faster finger propagation, and more branching at a higher injection pressure.\\
Finally also the spatial distance between the fingers depends on the injection pressure (see \ref{finger_k}). At low injection pressure the number of fingers is higher than at high injection pressure, as can be also seen in Fig. \ref{density}. In general the finger spatial frequency decreases in time after injection has started and fingers propagate through the cell. This can be shown by calculating the average of the characteristic spatial finger wavenumber in $x$ direction. First the powerspectrum $S_j$ of each horizontal line $j$ of the particle density is calculated. Taking the average of these power spectra results in a single power spectrum $\bar{S}$. From this average distribution of wave numbers the characteristic wave number $\langle k \rangle$ in $x$-direction is defined and calculated in the following way, using an average of $k$ with the power spectra as a weight:  
\begin{equation}
\label{avk}
\langle k \rangle =\frac{\sum_kk\bar{S}(k)}{\sum_k\bar{S}(k)}.
\end{equation} 
The results in Fig. \ref{finger_k} show a decrease of finger frequency in time. As a trend we notice that at higher injection pressure the finger frequency decreases faster than at low injection pressure. However the simulation at $P_I$=250 kPa differs from the other simulations. In this simulation we also observe a finger propagating directly along the right boundary in Fig. \ref{density}. Close to the wall this finger appears to propagate faster than the other fingers in this simulation. Because the simulation at $P_I$=250 kPa is the only simulation where this appears it also stands out in the plots for the average wave number Fig. \ref{finger_k}. This is presumably due to a finite size effect, and such outlier is frequently met in granular systems, which are known to present a large variability and sensitivity on details of the initial state. (see e.g. \cite{jlv3}). Otherwise for higher injection pressure does the finger frequency not only decreases faster but also drops to a lower value before the grains get compacted. \\
This coarsening of the finger frequency is the result of two mechanisms. First the pressure gradient between the finger tip and the outlet gets higher the closer the finger tip moves to the outlet. Assuming a linear pressure profile though the porous media the pressure would drop to zero on a shorter and shorter distance the closer the finger advances to the outlet. At the same time the gas also leaks into the side walls of the finger. This increases the pressure in the porous material around a finger. In the areas where this pressure increase takes place less advanced neighboring fingers would thus experience a lower pressure gradient. The speed of this fingers is thus reduced. This means the more a finger advances to the outlet the faster it moves. At the same time the pressure increase in the area around an advanced finger decreases the pressure gradient in front of less advanced fingers. This causes the less advanced fingers to propagate slower or to stop completely. This mechanism will result in a coarsening of the finger frequency. Further more we expect this mechanism to be active on a typical length scale which is comparable to the skin depth of the pressure profile. This mechanism also appears in the basic Saffman Taylor instability \cite{saff}.\\
A second mechanism that will account for a coarsening of the finger frequency is the compaction of the grains on the sides of a finger. During the propagation the finger width increases and branches at a 90 degree angle arise on the sides of fingers. This compacts the granular material on the sides of an advancing finger. How far this compaction propagates on the sides depends on the properties of the granular material and also on the finger width and how the side branches develop. Where this compaction has occur preceding fingers are slowed down or stopped. The size of the compaction front around the fingers sets a second length scale for the coarsening of the finger frequency.\\  
\section{Conclusions} 
The increase of the injection pressure primarily causes fingers to propagate faster through the granular packing. Fingers at high injection pressure also tend to be more branched and fracture-like than the fingers at low injection pressure. It was shown that the position of the fracture propagation in time increases with the square root of the injection pressure $\sqrt{P_I}$. 
Furthermore we discussed the observed coarsening of the characteristic spatial finger wavenumbers in terms of two mechanisms. A first mechanism that controls the coarsening arises from the fluid seepage into the granular media. Where the length scaled for this mechanism was argued to be of the size of the pressure skin depth. To further explain the coarsening a second mechanism causing the coarsening of the finger wavenumber was highlighted. This second mechanism introduces a length scale for the coarsening with the size of the compaction front in the granular material around a finger. Acknowledgments: We thank Gustavo S{\'a}nchez-Colina for help about the Spanish grammar.


\end{document}